    \documentclass[preprint,showpacs,preprintnumbers,superscriptaddress,amsmath,amssymb]{revtex4}
    \usepackage{cases}

    \usepackage{mathrsfs}
    \usepackage{amssymb}
    \usepackage[dvips]{graphicx}
    \usepackage{dcolumn}
    \usepackage{bm}
    \usepackage{subfigure}
    \usepackage{floatflt}
    \usepackage{float}
    \usepackage{rotating}
    \usepackage{multirow}
    \usepackage[bookmarksnumbered,bookmarksopen,colorlinks,citecolor=blue,linkcolor=blue]{hyperref} %

\begin{document}

    \preprint{00-000}

    \title{Mechanism of production of light complex particles in nucleon-induced reactions}

    \author{Dexian Wei}%
    \affiliation{College of Physics and Technology, Guangxi Normal University, Guilin, 541004, P. R. China}
    \author{Ning Wang}%
    \affiliation{College of Physics and Technology, Guangxi Normal University, Guilin, 541004, P. R. China}
    \author{Li Ou}
    \email{only.ouli@gmail.com}
    \affiliation{College of Physics and Technology, Guangxi Normal University, Guilin, 541004, P. R. China}

    \date{\today}

    \begin{abstract}

    The Improved Quantum Molecular Dynamics (ImQMD) model incorporated with the statistical decay model
    is successful in describing emission of nucleons in the intermediate energy spallation reactions,
    but not good enough in describing productions of light complex particles, i.e. $d$, $t$, $^3$He and $^4$He.
    To improve the description on emission of light complex particles,
    a phenomenological mechanism called surface coalescence and emission is introduced into ImQMD model:
    nucleon ready to escape from the compound nuclei can coalesce with the other nucleon(s) to form light complex particle
    and be emitted. With updated ImQMD model, the description on
    the experimental data of light complex particles produced in nucleon-induced reactions
     are great improved.
    \end{abstract}

    \maketitle



\section{Introduction}

\begin{figure*}[h]
    \includegraphics[width=12cm]{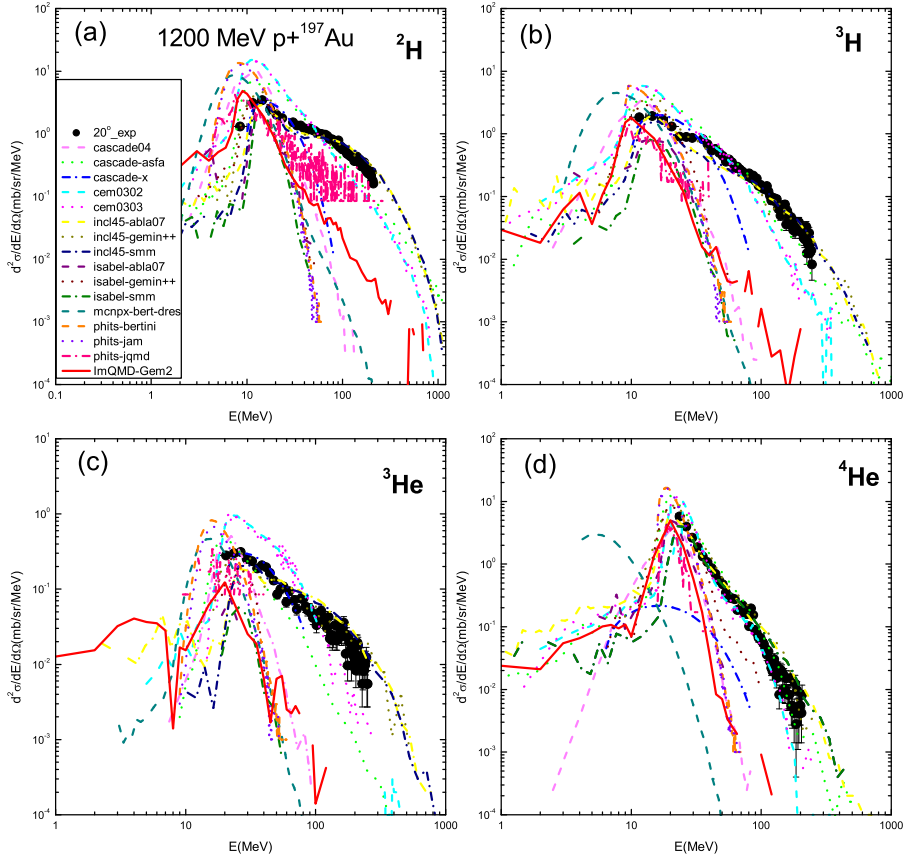}
    \caption{(Color online) Comparison of all IAEA benchmark models (including our original version of ImQMD model)
    to the experimental LCPs DDXs at 20$^{\circ}$ in 1200 MeV $p+^{197}$Au.
    Results of model calculations are taken from IAEA Benchmark of Spallation Models web site\cite{IAEAspa}.}
    \label{fig1}
    \end{figure*}

    Recently, because of the widely applications of the spallation reaction in various fields such as
    spallation neutron sources\cite{Bauer}, material science\cite{Gudow}, cosmography\cite{Smith},
    and Accelerator-Driven Subcritical Reactors (ADS) for nuclear waste transmutation\cite{Bowma,Abder}, etc,
    it recalls the attention in study on spallation reactions.
    A number of current and planned projects require a large amount of spallation reactions data\cite{Titar}.
    However, it is both physically and economically impossible to measure all necessary data\cite{NEA98}.
    So a theoretical model with powerful prediction ability is imperative.
    The International Atomic Energy Agency (IAEA) and the Abdus Salam International Center for Theoretical Physics
    have recently organized twice international workshop to make comparisons of spallation models and codes,
    including PHITS\cite{Iwase}, BUU\cite{Berts}, QMD\cite{Aiche}, JAM\cite{Nara1},
    JQMD\cite{Niit1}, INCL4\cite{Boud1}, ISABEL\cite{Yariv}, Bertini\cite{Bert1,Bert2}, Geant\cite{Agost},
    IQMD\cite{Hartnack}, RQMD\cite{Lehmann}, TQMD\cite{Puri} et al,
    by merged with various statistical decay models such as GEM\cite{Furihata00,Furihata02},
    GEMINI\cite{Chari}, ABLA\cite{Junghans} et al. The productions of the neutron, proton, pions and isotopes
    can be overall described well by the most of given models\cite{Titar}.
    But the data for the light complex particles (LCPs), i.e. $d$, $t$, $^3$He, and $^4$He,
    can not be reproduced well by the most of models\cite{Blide}, as shown in Fig. \ref{fig1}
    the comparison of all IAEA benchmark models (including our original version of Improved Quantum Molecular Dynamics (ImQMD) Model)
    to the experimental LCPs double differential cross sections (DDXs) at 20$^{\circ}$ in 1200 MeV $p+^{197}$Au.
    Only models which have a specific mechanism to emit energetic clusters,
    such as coalescence during the intranuclear cascade stage or pre-equilibrium emission of composite nuclei,
    can reproduce the high-energy tail of LCPs DDXs\cite{Lera1}.
    The information of yield of hydrogen and helium element is quite important for design of nuclear project
    including target and shield.
    Therefore, the description of model on LCPs emission should be improved to satisfy the request of
    the applications\cite{Lera1,DAVID}.

    It is well known that spallation reaction is usually described by three-step processes,
    i.e. the dynamical non-equilibrium reaction process leading to the emission of fast particles
    and an excited residue, followed by pre-equilibrium emission, and by the decay of the residue.
    The first process can be described by microscopic transport theory models,
    the pre-equilibrium is usually optional in different approaches,
    and the last one can be described by a statistical decay model.
    By applying the Improved Quantum Molecular Dynamics (ImQMD) Model merged statistical decay model,
    a series of studies on the proton-induced spallation reactions
    at intermediate energies have been made\cite{Liou1, Liou2, Liou3}.
    Nuclear data including neutron DDXs, proton DDXs,
    mass, charge and isotope distributions can be overall reproduced quite well.
    However, the yields of LCPs with high kinetic energy are underestimated in the previous ImQMD model,
    as shown in Fig. \ref{fig1},
    because the description of LCPs emission in pre-equilibrium process is absent.
    The motivation of this work is to improve the description on LCPs in ImQMD model.
    The pre-equilibrium LCP emission has been described reasonably well by implementation of
    a surface coalescence mechanism in INC and INCL model in nucleon-induced reactions\cite{Letou,Herba,Budza}.
    The same method was introduced into JQMD code by Watanabe and the great improvement has been achieved\cite{Watan}.
    In the present work, we also introduce such a surface coalescence mechanism into the ImQMD05 model
    to make systemic studies on LCPs emission in various nucleon-induced reactions.

    The paper is organized as follows. In the next section, we
    make a brief introduction on ImQMD05 model and how to introduce the surface coalescence model into ImQMD05 simulation.
    In Sec. III, we give the calculation results and make some discussion. Finally, a summary is given in Sec. IV.

    \section{Model}
    \subsection{ImQMD05 model}
    ImQMD model is an extend version of QMD\cite{Aiche} and IQMD\cite{Hartnack}.
    A detailed description of the ImQMD model and its version ImQMD05 and their
    applications can be found in Refs. \cite{Zh05,Zh06,Liou1,Liou2,Liou3,Liou4}.
    Within the ImQMD05 model,
    the nucleon is represented by a Gaussian wave packet
    \begin{eqnarray}\label{Eqnarry1}
        \phi_{i}(\bm{r})=\frac{1}{(2\pi\sigma_{r}^{2})^{3/4}}
        \exp\left[-\frac{(\bm{r}-\bm{r}_{i})^{2}}{4\sigma_{r}^{2}}
        +\frac{i}{\hbar}\bm{r}\cdot\bm{p}_{i}\right],
    \end{eqnarray}
    Where $\bm{r}_{i}$ and $\bm{p}_{i}$ are the centers of wave packet of the $i$th nucleon in the coordinate
    and momentum space, respectively.
    The wave-packet width $\sigma_{r}$ in the QMD model can be regarded as a quantity
    having relations with the interaction range of a particle.
    For finite systems, particles should be localized in a finite size corresponding to the size of the system,
    and thus the wave-packet width should have some relations with the range of the mean field
    which binds particles together and should change with the time evolvement.
    In practice, one already notes that the value of the wave-packet width
    affects the calculation results obviously, so that one usually makes an adjustment to a certain
    extent\cite{Hartnack,Wang,Gautam}.
     According to the results from other group\cite{Niit1} and our previous studies,
    a tradition value of $\sigma_{r}^{2}=2.0$ fm$^{2}$ is appropriate for spallation reactions.
    And $\sigma_{p}$ is fixed by
    uncertainty relation $\sigma_{r}\cdot\sigma_{p}=\frac{\hbar}{2}$.
    The one-body phase space distribution function is obtained by the
    Wigner transform of the wave function which reads
    \begin{eqnarray}\label{Eqnarry2}
    f(\bm{r},\bm{p})=\sum_{i=1}^{A}\frac{1}{(\pi\hbar)^3}
        \exp\left[-\frac{(\bm{r}-\bm{r}_i)^2}{2\sigma_{r}^{2}}\right]
        \exp\left[-\frac{(\bm{p}-\bm{p}_i)^2}{2\sigma_{p}^{2}}\right].
    \end{eqnarray}
    Nucleons in a system move under the mean-field, and the time evolution of $\bm{r}_{i}$ and
    $\bm{p}_{i}$ is governed by Hamiltonian equations of motion
    \begin{eqnarray}\label{Eqnarry3}
    \dot{\bm{r}}_{i}=\frac{\partial H}{\partial\bm{p}_{i}}, \;\; \;
    \dot{\bm{p}}_{i}=-\frac{\partial H}{\partial\bm{r}_{i}},
    \end{eqnarray}
    where
    \begin{eqnarray}\label{Eqnarry4}
    H=T+U_{\rm{loc}}+U_{\rm{Coul}},
    \end{eqnarray}
    here $T$ and $U_{\rm{Coul}}$ is the kinetic energy and the Coulomb potential energy, respectively.
    And the local potential energy ${\it{U_{\rm{loc}}}}=\int V_{\rm{loc}}[\rho(\bm{r})] d \bm{r}$, where
    $V_{\rm{loc}}[\rho(\bm{r})]$ is the nuclear potential energy density functional
    including the full Skyrme potential energy density with just the spin-orbit term omitted, which reads
    \begin{align}\label{Eqnarry5}
    V_{\rm{loc}}=&\frac{\alpha}{2}\frac{\rho ^{2}}{\rho _{0}}+\frac{\beta }{\eta +1}%
    \frac{\rho ^{\eta +1}}{\rho _{0}^{\eta }}+\frac{g_{\rm{sur}}}{2\rho _{0}}\left(\nabla \rho \right)^{2}
    +\frac{g_{\rm{sur,iso}}}{\rho_{0}}[\nabla(\rho_{n}-\rho_{p})]^{2} \nonumber\\
    &+(A\rho^{2}+B\rho^{\eta+1}+C\rho^{8/3})\delta^{2}+g_{\rho\tau}\frac{\rho^{8/3}}{\rho_{0}^{5/3}},
    \end{align}
    The parameters in Eq. \eqref{Eqnarry5} are fully determined by Skyrme interactions.
    where $\rho$, $\rho_{n}$, $\rho_{p}$ are the nucleon, neutron, and
    proton density, $\delta = (\rho_{n}-\rho_{p})/(\rho_{n}+\rho_{p})$
    is the isospin asymmetry. The first two terms in
    Eq.~(\ref{Eqnarry5}) are the iso-scalar bulk potential energy part, the
    third term is the isospin independent surface energy term, the
    forth term is the surface symmetry energy term and the fifth term
    is the bulk symmetry potential energy term. The last term, called the
    $\rho\tau$ term, is obtained from the $\rho\tau$ term of the
    Skyrme potential energy density functional by applying the
    Thomas-Fermi approximation to the kinetic energy density $\tau$.
    However, the strength of this term $g_{\rho\tau}$ is rather small
    compared with other iso-scalar terms. The coefficients in
    Eq.~(\ref{Eqnarry5}) are therefore directly related to the
    standard Skyrme interaction parameters as
    \begin{align}\label{Eqnarry6}
    \frac{\alpha}{2}&=\frac{3}{8}t_{0}\rho_{0},~~
    \frac{\beta}{\eta+1}=\frac{1}{16}t_{3}\rho_{0}^{\eta},\nonumber\\
    \frac{g_{\texttt{sur}}}{2}&=\frac{1}{64}(9t_{1}-5t_{2}-4x_{2}t_{2})\rho_{0},\\
    \frac{g_{\texttt{sur,iso}}}{2}&=-\frac{1}{64}[3t_{1}(2x_{1}+1)+t_{2}(2x_{2}+1)]\rho_{0}.\nonumber
    \end{align}
    And the $A$, $B$ and $C$ in the volume symmetry potential energy term are
    also given by the Skyrme interaction parameters,
    \begin{align}\label{Eqnarry7}
    A&=-\frac{t_{0}}{4}(x_{0}+1/2),\nonumber\\
    B&=-\frac{t_{3}}{24}(x_{3}+1/2),\\
    C&=-\frac{1}{24}\left(\frac{3\pi^{2}}{2}\right)^{2/3}\Theta_{\texttt{sym}},\nonumber
    \end{align}
    where $\Theta_{\texttt{sym}}=3t_{1}x_{1}-t_{2}(4+5x_{2})$.  The
    $g_{\rho\tau}$ is determined by
    \begin{equation}\label{Eqnarry8}
    g_{\rho\tau}=\frac{3}{80}[3t_{1}+(5+4x_{2})t_{2}]\left(\frac{3\pi}{2}\right)^{2/3}\rho_{0}^{5/3}.
    \end{equation}
    The $t_{0},~t_{1},~t_{2},~t_{3}$ and $x_{0},~x_{1},~x_{2},~x_{3}$ in
    expressions (\ref{Eqnarry6})-(\ref{Eqnarry8}) are the standard parameters of Skyrme force.
    With SkP parameter set, ImQMD model has been used to describe
    the heavy-ion collision\cite{Zh05,Zh06} and spallation reaction\cite{Liou1,Liou2,Liou3} successfully
    in our previous studies, so in this work we still choose SkP parameter set\cite{Dob84} in the calculations.

    In the collision term, the free isospin dependent nucleon-nucleon
    scattering cross sections\cite{Cugnon96} are adopted in the calculations.
    The Pauli principle is considered carefully in the collision term. Neutrons and protons
    are treated as different particles. According to the uncertainty principle,
    the final states of two particles are required to satisfy the relation
    \begin{eqnarray}
    \frac{4\pi}{3} r_{ij}^{3} \cdot \frac{4\pi}{3} p_{ij}^{3} \geq
    \frac{h^{3}}{8}.
    \end{eqnarray}
    The $r_{ij}$ and $p_{ij}$ are the distances between the centers of wave packets of two nucleons in
    coordinate and momentum space.
    The possibility being blocked at the final state $i$ and $j$ for each of the two scattering
    nucleons is calculated by
    \begin{eqnarray}
    P_{\rm{block}}=1-(1-P_{i})(1-P_{j}).
    \end{eqnarray}
    The $P_{i}$ and $P_{i}$ are the probabilities for state $i$ and $j$ being occupied, which can be calculated by
    \begin{eqnarray}
    P_{i}=\sum_{k,k\neq i}^{A} \frac{1}{(\pi \hbar)^{3}}
    \exp\left[-\frac{(\bm{r}_{i}-\bm{r}_{k})^{2}}{2\sigma_{r}^{2}}\right]
    \exp\left[-\frac{(\bm{p}_{i}-\bm{p}_{k})^{2}}{2\sigma_{p}^{2}}\right].
    \end{eqnarray}

    At the end of the ImQMD05 calculations, clusters are recognized
    by a minimum spanning tree (MST) algorithm\cite{Aiche} widely used in the QMD calculations.
    In this method, the nucleons with relative momenta smaller than $P_{\rm{c}}$ and relative distances smaller than $R_{\rm{c}}$
    are coalesced into the same cluster.
    This approaches has been quite successful in explaining certain fragmentation observables
    such as charge distributions of the emitted particles, intermediate mass fragment multiplicities\cite{Aiche,Zh05,Zh06}.
    But the MST method fails to describe other detail in the production of nucleons
    and light charged particles\cite{Aiche,Zh05,Zh12,Li13,Nebauer}.
    For example, the yields of $Z=1$ particles are overestimated, while the yields of $Z=2$ particles are underestimated
    partly due to the strong binding of $\alpha$ particles.
    In this work, $R_{\rm{c}}$ = 4.5 fm and $P_{\rm{c}}$ = 250 MeV/$c$ are adopted.
    Then the total energy of each excited cluster is calculated in its rest frame and its
    excitation energy are obtained by subtracting the corresponding ground state energy from
    the total energy of the excited cluster.
    The information of excited cluster are input into the statistical decay model GEM2\cite{Furihata00,Furihata02}
    to perform statistical decay stage calculations.

\subsection{Mechanism of surface coalescence}

    The study of the pre-equilibrium nuclear reactions has been an active topic since the pioneering work
    by Goldberger\cite{Goldb} and Metropolis\cite{Metr2} based on the cascade model.
    But there is still not a well understood mechanism of composite particles emission before equilibrium
    because of the complexity of the pre-equilibrium process.
    One of concepts dealing with LCPs which momenta are close to that of the incident projectile is
    ``pick-up'' or ``knock-out'' reaction\cite{Hodgs, Barlo}.
    In this concept the complex particles to be generated
    through successive elementary reactions as $p+n\longrightarrow d+\pi^0$, $d+p\longrightarrow ^3$He$+\pi^0$, etc.
    Another method to deal with LCPs contributing at any angle
    is coalescence of nucleons with the involved particles close enough in phase space.
    This non-direct mechanism for composite particles production was first discussed by Butler and Pearson\cite{Butler62}.
    And some modifications have been made by Nagle\cite{Nagle} and Mattiello et al\cite{Matti} to
    investigate the possible formation of composite particles during the INC stage.
    Then the surface coalescence mechanism has been incorporated in various model, including INCL\cite{Letou,Herba,Budza},
    QMD\cite{Watan}, and coalescence exciton model\cite{Iwamo}, to investigate the LCPs emission in the pre-equilibrium process.

    With the similar idea, the surface coalescence mechanism is introduced into ImQMD05 model
    as follows: After incident nucleon touching the target nuclei to form the compound nuclei,
    we define a sphere core with radius $R_0$ surrounding by a surface with width $D_0$.
    At each time step, any fast nucleon passing the surface region to leave
    the compound system is taken as leading nucleon.
    An inspection is made over all other regular nucleons in order to check whether
    there are one or several nucleons close enough in phase space to allow the formation of a stable composite-particle.
    If the the phase space condition is satisfied, the candidate LCP will be constructed
    staring from the leading nucleon, by finding a second, then a third, etc, nucleon satisfying the following condition
    \begin{equation}
    R_{im}\times P_{im}\leq h_0 \hspace{0.5cm}
    {\rm{with}} \hspace{0.5cm} R_{im}\geq 1~\rm{fm},
    \end{equation}
    Where the $R_{im}$ and $P_{im}$ are the Jacobian coordinates of the $i$th nucleon,
    i.e., the relative spatial and momentum coordinates of the considered $i$th nucleon with respect to the subgroup cluster $m$.
    The value of $h_0$ is adjustable by fitting the experimental data.
    And the condition $R_{im}\geq 1~\rm{fm}$ gets rid of the unreal LCPs
    constituted by the nucleons being too close to each other,
    due to the repulsion of nucleon-nucleon interaction in short distance\cite{Letou}.
    In present work, the following LCPs are considered: $d$, $t$, $^3$He, $^4$He.
    By the method of LCPs to be constructed, candidate nucleon belongs to a heavier LCP
    also belongs to a lighter LCP. So LCPs are checked to be emitted according to the priority list:
    $^4$He$>^3$He$>t>d$, say the heavier LCP is first tested for emission,
    as the same order as that in Ref.\cite{Letou,Boud2,Watan}.
    Finally the candidate LCP can be emitted or not depends on whether its kinetic energy
    is high enough to tunnel through the Coulomb barrier. The kinetic energy of the LCPs can be calculated as
    \begin{equation}
    E_{\rm{lcp}}=\sum_{i=1}^{A_{\rm{lcp}}} (E_i+V_i)+B_{\rm{lcp}}
    \end{equation}
    Where the $E_i$ and $V_i$ are the kinetic energy and the potential energy of
    the $i$th constituent nucleon, respectively, $A_{\rm{lcp}}$ and $B_{\rm{lcp}}$ are the mass number
    and the binding energy of the LCP, respectively. If all conditions are meet,
    the LCP is emitted in the direction of its c.m. momentum.
    Otherwise, all nucleons in the ``LCP'' are set free and become available again in the nucleus and in the ImQMD process,
    the leading nucleon is emitted as a free nucleon.
    In this coalescence model, composite-particles are thus not allowed being formed in
    the interior of the nucleus but only in the surface layer.
    It is reasonable according to the knowledge from nuclear structure and reactions.
    For each leading nucleon, the LCP formation and emission are tested in the priority list.
    At each time step, the same test is repeated until the end of ImQMD simulation.

\section{Results and discussion}

    The nucleon-induced reactions on various targets with incident energy from 62 MeV to 1.2 GeV
    are simulated by ImQMD05 with the surface coalescence mechanism introduced.
    Firstly, systemic analysis on spectra and DDXs of $p$, $d$, $t$, $^3$He, $^4$He are performed
    to fix the parameters in the surface coalescence model.
    The spectrum and DDX of emitting particle are calculated respectively by
    \begin{align}
    \frac{d \sigma}{dE}&=\sum\limits_{i=1}^{i_{\rm{max}}}2 \pi b_{i}\Delta b f(E,b_{i}),\\
    \frac{d^{2} \sigma}{d\Omega dE}&=\sum\limits_{i=1}^{i_{\rm{max}}}2 \pi b_{i}\Delta b f(E,\Omega,b_{i}),
    \end{align}
    where $f(E,b_{i})$ is the possibility of a particle emitted with kinetic energy $E$,
    and $f(E,\Omega,b_{i})$ is the possibility of a particle with kinetic energy $E$ emitted into solid angle $\Omega$,
    under impact parameter $b_{i}$. And the maximum impact parameter
    $b_{\rm{max}}$=5.0, 5.5, 7.0, 7.0, 8.0, 9.0, 9.5, 9.5 fm for the reactions of nucleon-induced targets
    $^{16}$O, $^{27}$Al, $^{56}$Fe, $^{58}$Ni, $^{120}$Sn, $^{197}$Au, $^{208}$Pb, $^{209}$Bi,
    respectively. In the calculations, for each impact parameter, 50,000 events are performed.
    The freeze-out time is different for various targets, depending on target mass.
    In order to save CPU time and get the best results, the switching time from the ImQMD05
    to GEM2 is taken as short as possible for various targets.
    It is taken as 100 fm/$c$ for light nucleus ($A\leq27$),
    125 fm/$c$ for intermediate nucleus $(27\leq A\leq58)$,
    and 150 fm/$c$ for heavy nucleus $(A>58)$.
    Although the dynamical interaction is not totally complete,
    after the switching time only some free nucleons with low energy emit from the residua,
    which can be describe by the evaporation of nucleons in the decay stage.

\begin{figure*}[h]
    \includegraphics[height=7cm]{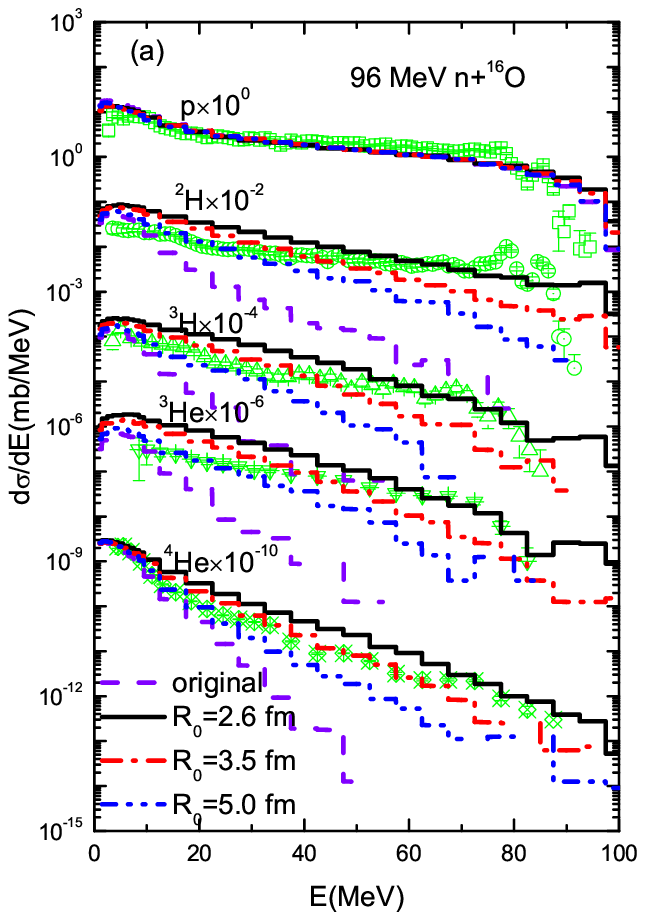}    
    \includegraphics[height=7cm]{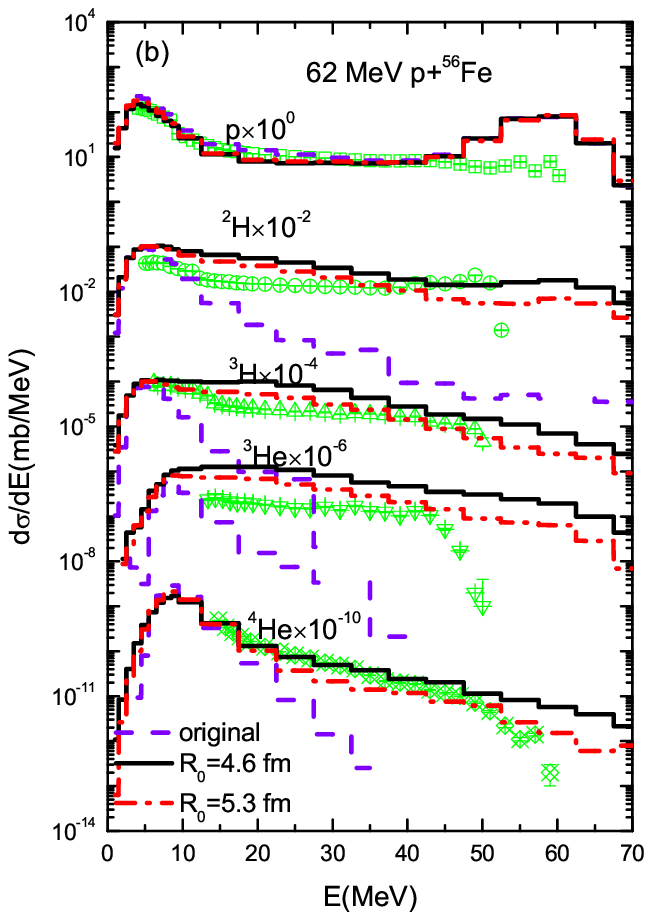}
    \includegraphics[height=7cm]{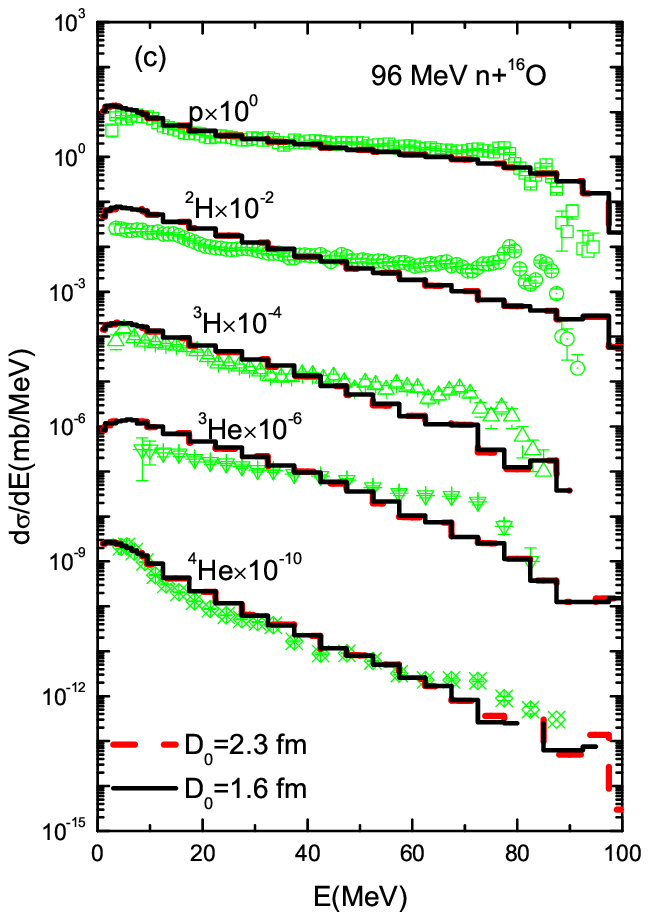}
    \caption{(Color online) Energy spectra of light complex particles calculated with various parameters
    in the reaction $n+^{16}$O at 96 MeV (in subfigures (a), (c), experimental data are taken from Ref.\cite{Tippa}),
    and $p+^{56}$Fe at 62 MeV (in subfigures (b), experimental data are taken from Ref. \cite{Bert4}).
     The spectra for various LCPs are multiplied by various factors noted in the figures.}
    \label{fig2}
    \end{figure*}

    There are some adjustable parameters in the surface coalescence model to be fixed.
    Firstly, $R_0$ is examined with $D_0=2.3$ fm and $h_0=200$ fm MeV/$c$ referenced to Ref.\cite{Watan}.
    In Fig. \ref{fig2}, the spectra of all LCPs emitted in the reaction $n+^{16}$O at 96 MeV (in subfigure (a)),
    $p+^{56}$Fe at 62 MeV (in subfigure (b)) calculated with various $R_0$ are compared to the experimental data.
    The results calculated by original ImQMD05 model without the surface coalescence model
    are also shown as dashed lines in Fig. \ref{fig2}(a) and (b).
    As a rule in this article and for the sake of clarity,
    cross-sections are displayed after multiplication by 10$^0$, 10$^{-2}$, 10$^{-4}$, etc.,
    as noted in the figures.
    One can see that there is a great enhancement of yield of LCPs in the high energy tail
    with the surface coalescence mechanism introduced.
    The value of $R_0$ affects the LCPs yields obviously, because it determines the possibility of
    the nucleon to be a leading nucleon.
    Smaller $R_0$ makes more nucleon to have opportunity to be leading nucleon,
    then the yields of LCPs increase.
    By systematic studies, it is found that calculations with $R_0 \simeq 1.4A^{1/3}~{\rm{fm}}$
    ($A$ is the mass number of target) can describe overall experimental data quite well.
    Then effect of $D_0$ on LCPs emission is investigated with $D_0=1.6, 2.3$ fm
    and $h_0=200$ fm MeV/$c$, $R_0=3.5$ fm,  in $n+^{16}$O at 96 MeV reaction,
    as shown in Fig. \ref{fig2}(c).
    One can see that there is not obvious difference between two results.
    It is easy to understand this phenomenon. Although larger $D_0$ involves more nucleon far away from sphere core
    to be leading nucleon, it is difficult for those leading nucleon to ``pick-up''
    nucleon(s) from sphere core to formed LCPs, because their relative spatial coordinate is too large.
    So the yields of LCPs are not sensitive to $D_0$.
    To be sure each nucleon in surface has more probability to be a leading nucleon,
    thicker surface with $D_0=2.3$ fm is still adopted in the following calculations.

\begin{figure*}[h]
    \includegraphics[height=14cm]{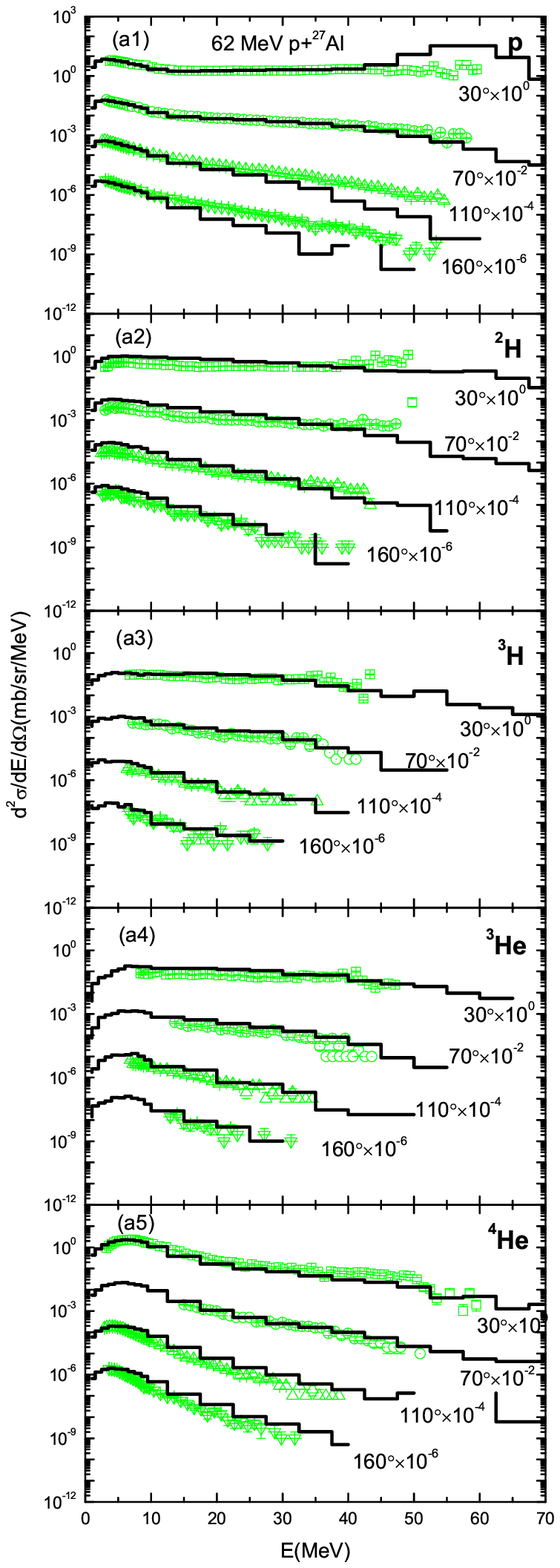}
    \includegraphics[height=14cm]{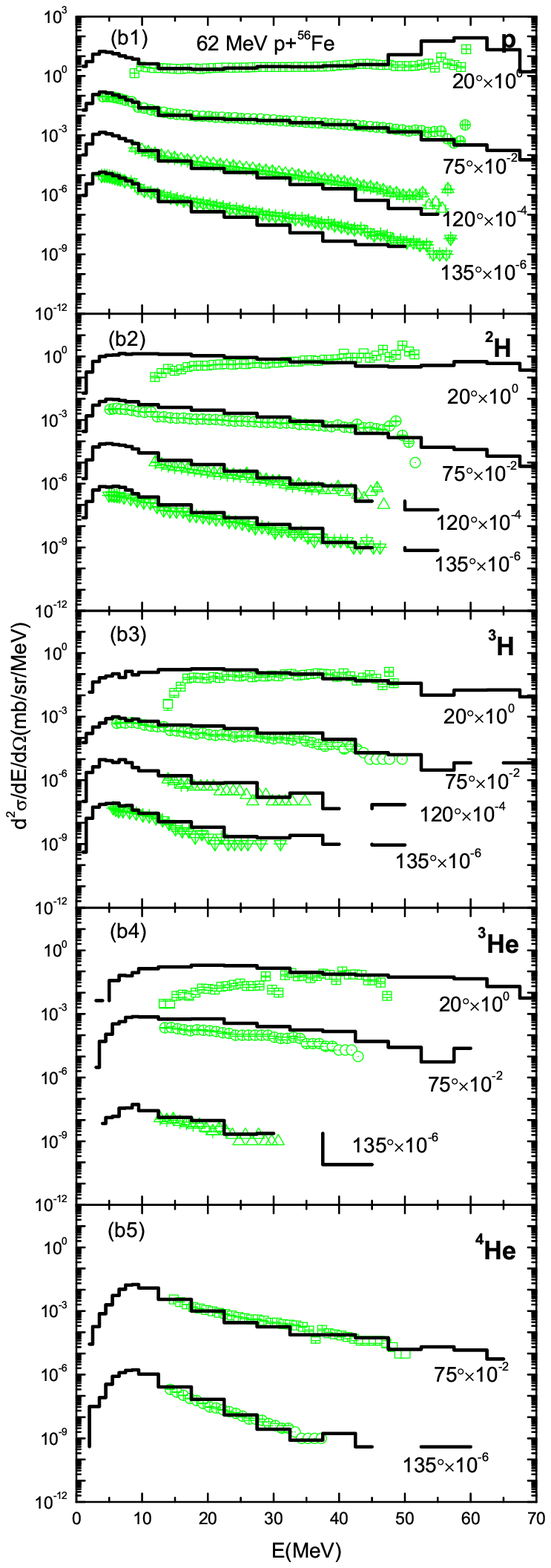}
    \includegraphics[height=14cm]{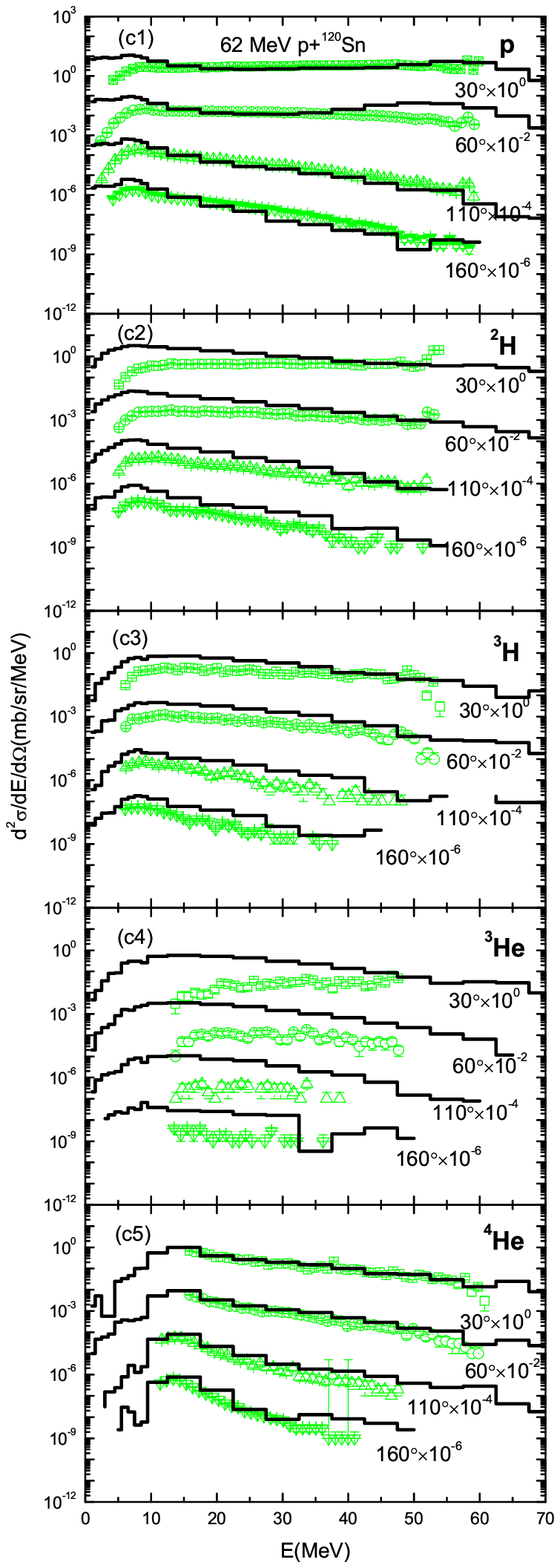}
    \caption{(Color online) Comparisons between calculations and experimental data of DDXs for light charged particles
    in the reaction $p+^{27}$Al (left pane), $p+^{56}$Fe (middle pane) and $p+^{120}$Sn (right pane) at 62 MeV.
     Experimental data are taken from Ref.\cite{Bert4}.}
    \label{fig3}
    \end{figure*}

\begin{figure*}[h]
    \includegraphics[height=5cm]{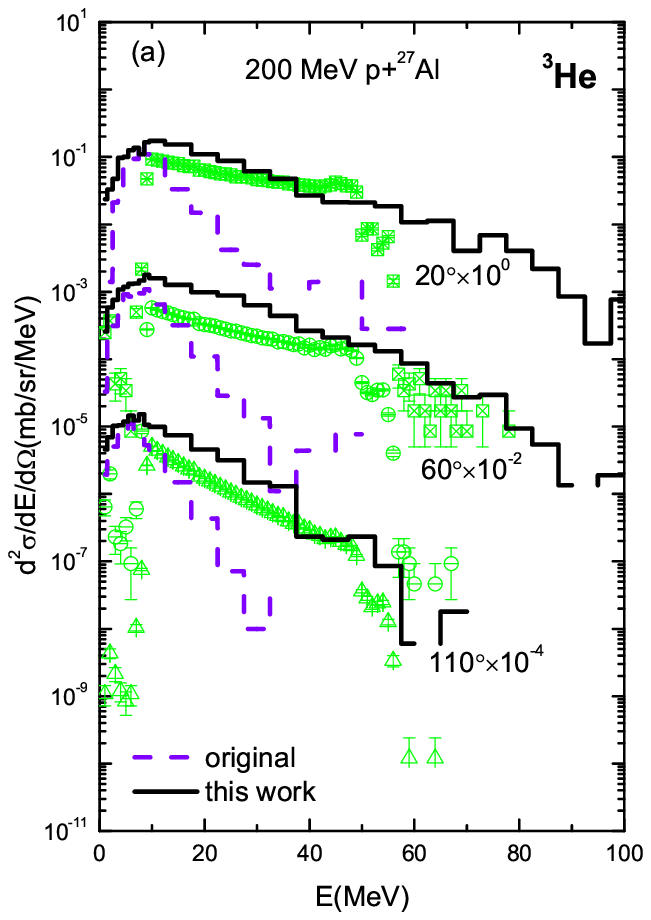}
    \includegraphics[height=5cm]{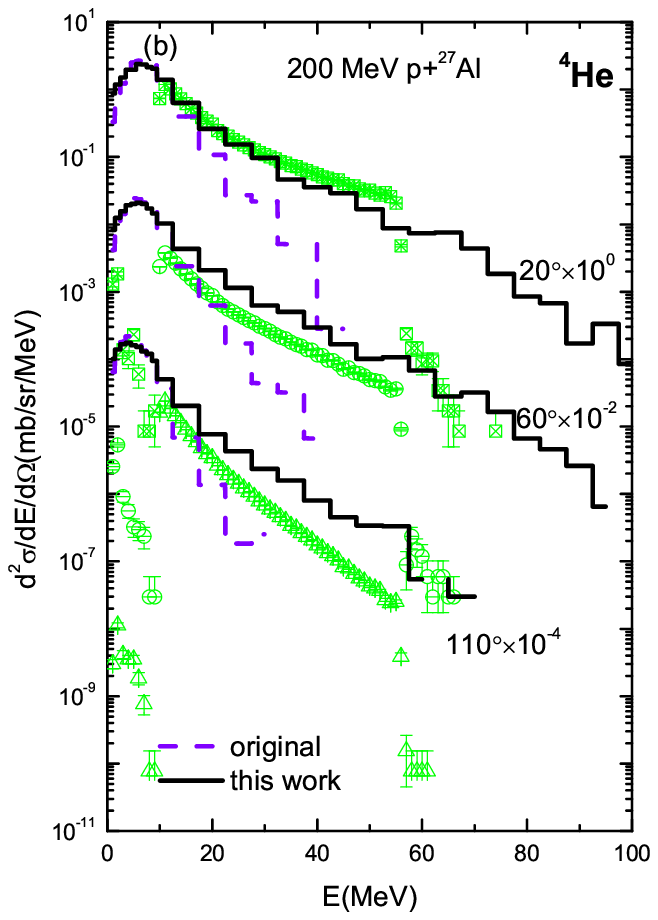}
    \includegraphics[height=5cm]{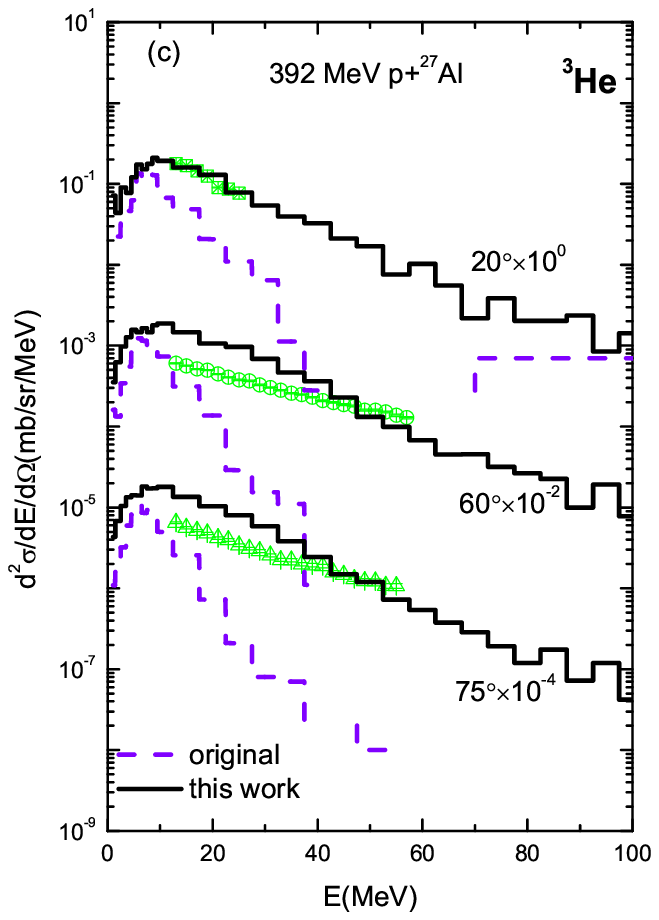}
    \includegraphics[height=5cm]{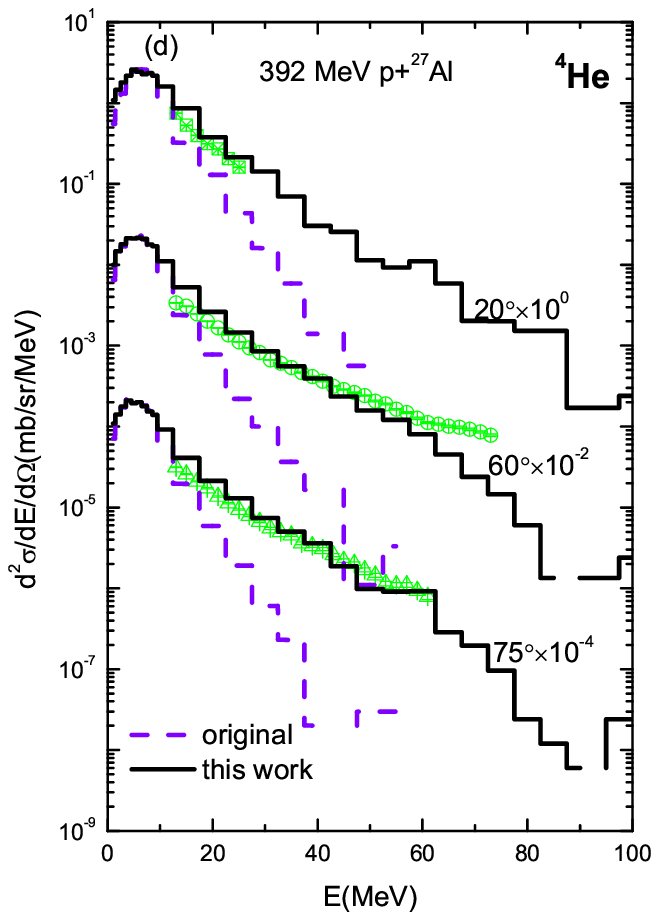}
    \caption{(Color online) Calculated DDXs of light charged particles
    produced in the reaction $p+^{27}$Al at 200, and 392 MeV.
    Experimental data are taken from Ref.\cite{Mach1} for 200 MeV $p+^{27}$Al,
    Ref.\cite{Uozumi2007} for 392 MeV $p+^{27}$Al, respectively.}
    \label{fig4}
    \end{figure*}

\begin{figure*}[h]
    \includegraphics[height=5cm]{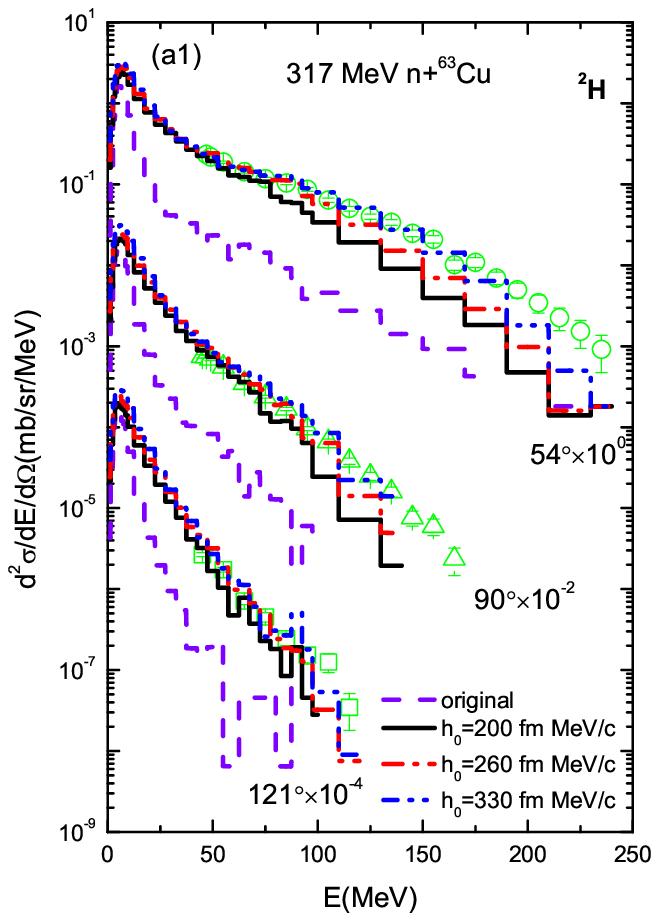}
    \includegraphics[height=5cm]{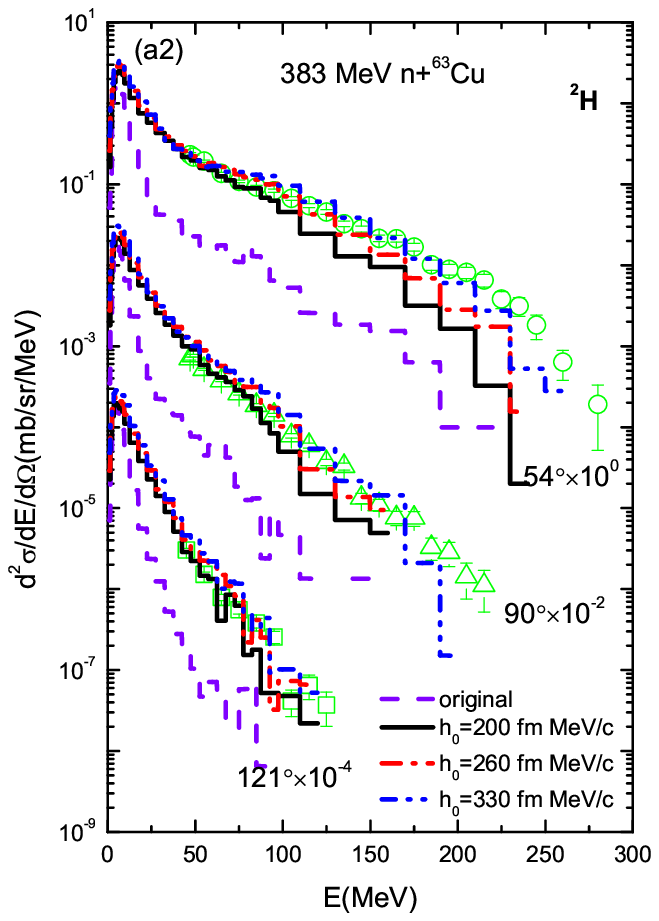}
    \includegraphics[height=5cm]{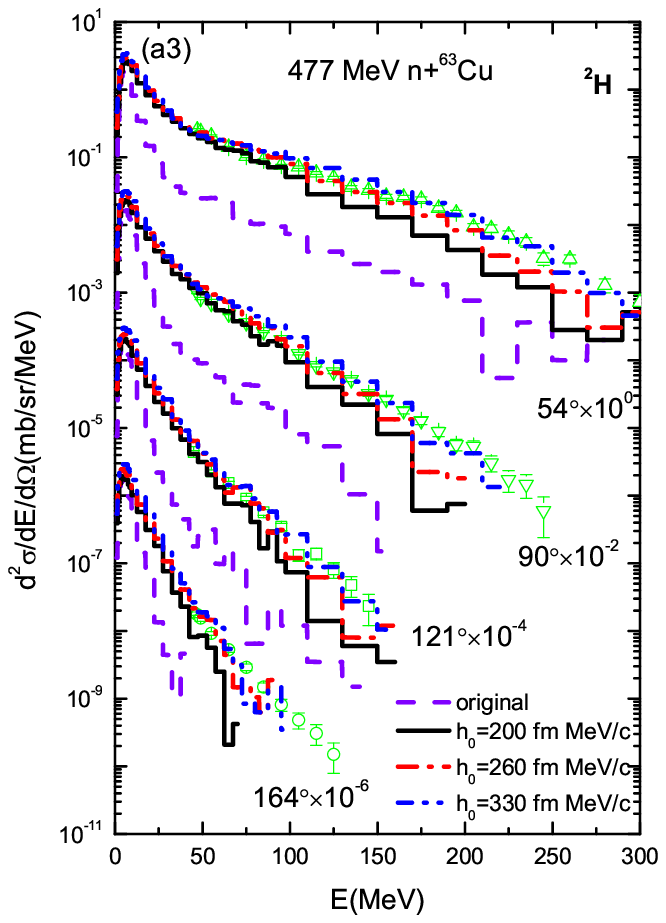}
    \includegraphics[height=5cm]{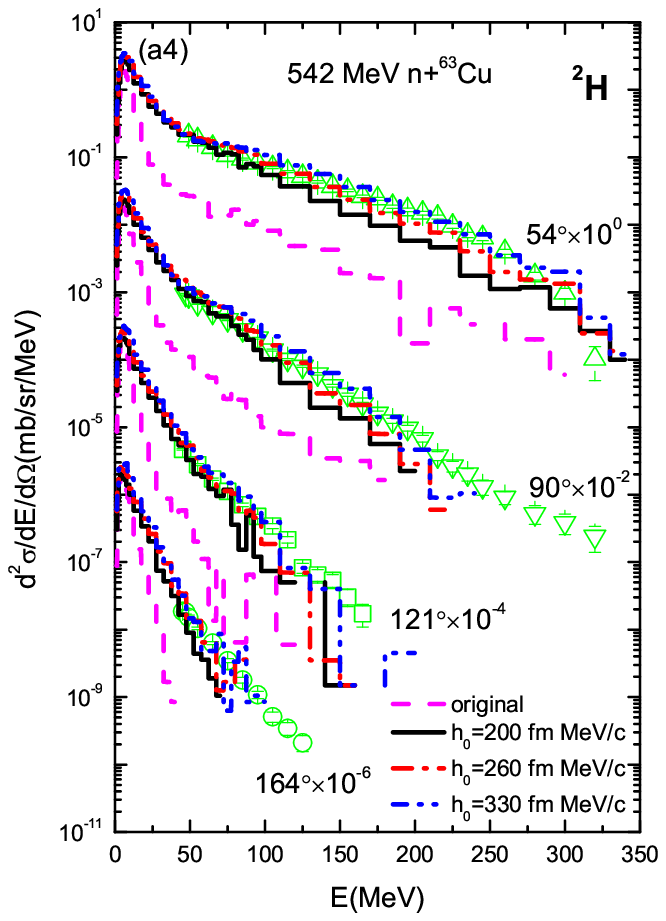}\\
    \includegraphics[height=5cm]{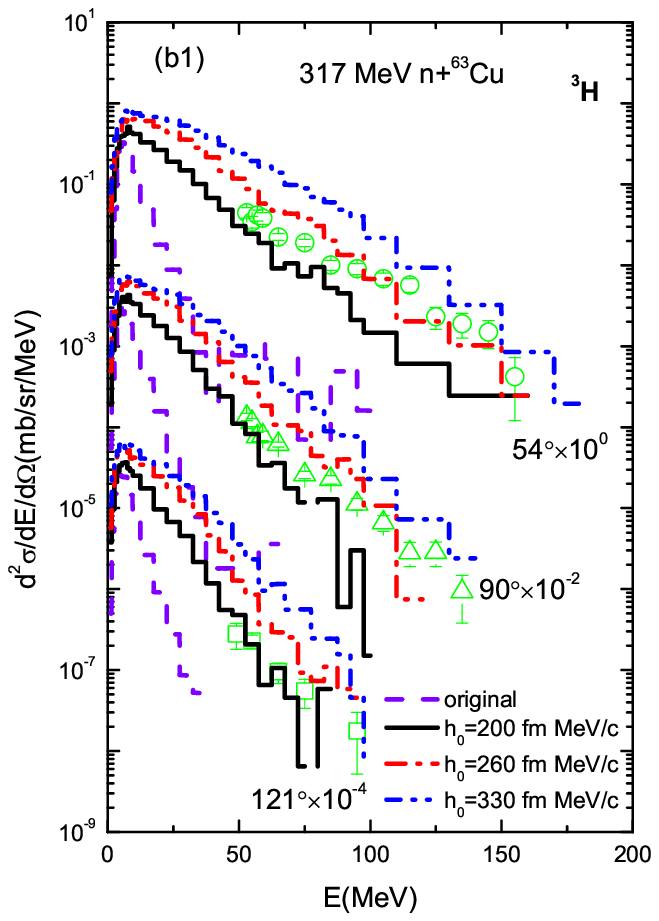}
    \includegraphics[height=5cm]{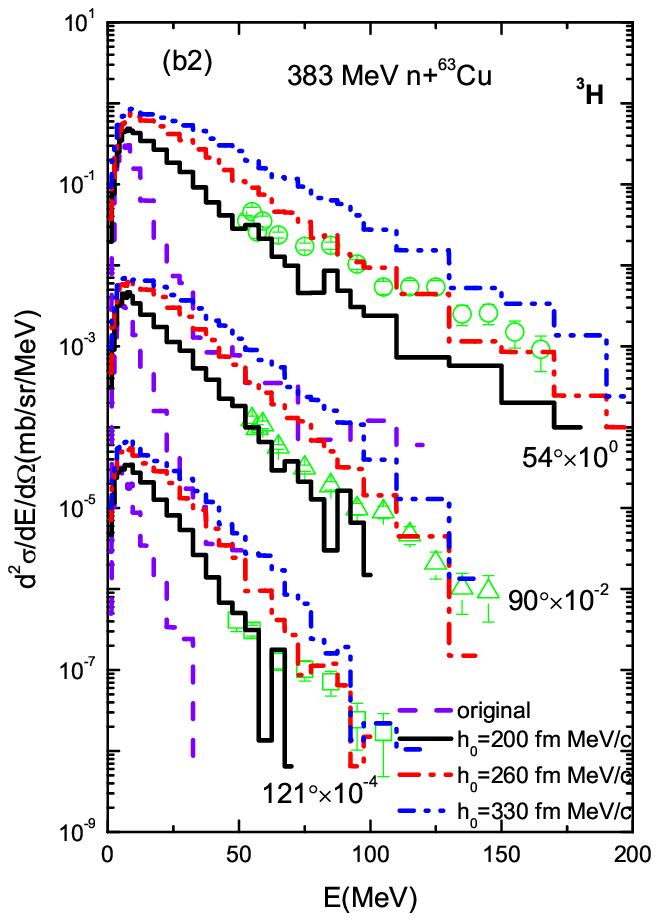}
    \includegraphics[height=5cm]{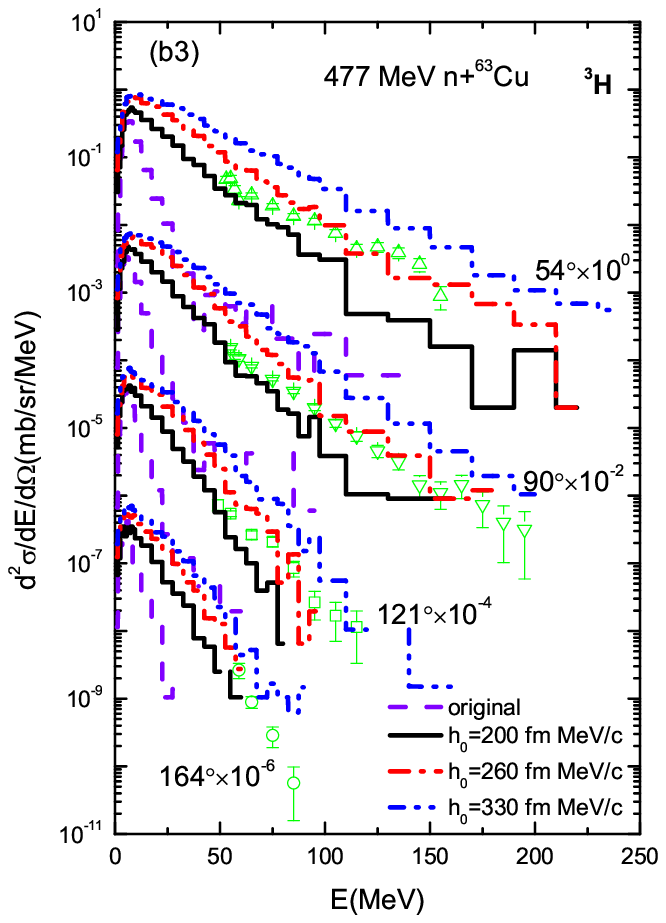}
    \includegraphics[height=5cm]{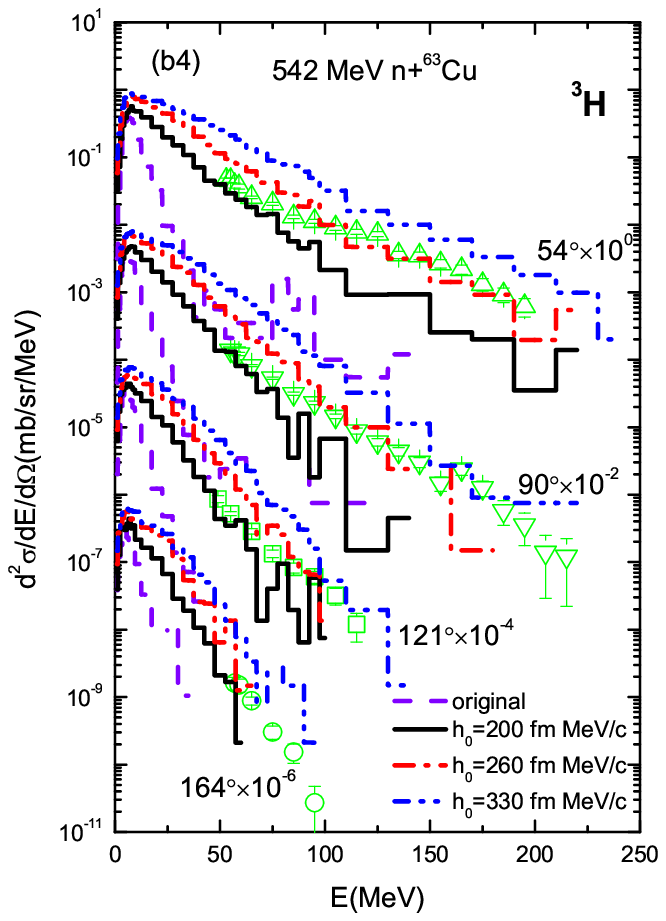}
    \caption{(Color online) Calculated DDXs of light charged particles
    produced in the reaction $n+^{63}$Cu at 317, 383, 477 and 542 MeV, respectively,
    with $h_0$=200, 260, and 330 MeV fm/$c$ is adopted.
    Experimental data are taken from Ref.\cite{Franz}.}
    \label{fig5}
    \end{figure*}

    Next, the parameter $h_0$ is investigated.
    With $h_0=200$ fm MeV/$c$, $R_0=1.4A^{1/3}$ fm, and $D_0=2.3$ fm,
    62 MeV proton bombarding $^{27}$Al, $^{56}$Fe, and $^{120}$Sn,
    and 200, and 392 MeV proton bombarding $^{27}$Al at are simulated.
    Figures \ref{fig3} and \ref{fig4} shows comparisons between calculated DDXs and experimental data\cite{Bert4,Mach1,Uozumi2007}
    for light charged particles. One can see that, with these parameters,
    the calculations are in good agreement with the experimental data,
    except DDXs at low energy range in forward angle for heavy targets, i.e. $^{56}$Fe and $^{120}$Sn.
    Then the reactions with higher energies are studies.
    And it is found that, $h_0=200$ MeV fm/$c$ is not good enough to describe the data.
    Figure \ref{fig5} shows the calculated $^2$H, $^3$H
    produced in the reaction  $n+^{63}$Cu at 317, 383, 477 and 542 MeV, respectively,
    with $h_0$=200, 260, and 330 MeV fm/$c$ is adopted.
    From the figure, one can see that $h_0$ affects DDX obviously. with $h_0$=200 MeV fm/$c$,
    calculated DDXs of LCPs for the case of 317 MeV are in good agreement with the experimental data.
    But for the case of 383, 447 and 542 MeV, calculations with $h_0$=200 MeV fm/$c$ underestimate the experimental data.
    And these deviations should be corrected by adjusting $h_0$, but not by adjusting $R_0$, and $D_0$.
    With $h_0$=260, 330 MeV fm/$c$, the cross sections are enhanced obviously,
    because with larger $h_0$, the limit for nucleons involved into a LCP becomes looser,
    the opportunity of leading nucleon coalescing with other nucleons to form LCP is enhanced.
    When $h_0$ is increase to 260 MeV fm/$c$, the experimental data can be reproduced quite well.
    So $h_0$ depends on the incident energy, according to the calculation results,
     in present study, $h_0$ is set to be
    \begin{subnumcases}
    {h_0=}
    200~{\rm MeV~fm}/c, &$E_{\rm{lab}}\leq 300$ MeV,\nonumber\\
    260~{\rm MeV~fm}/c, &$300~{\rm{MeV}} < E_{\rm{lab}}\leq 500$ MeV,\nonumber\\
    330~{\rm MeV~fm}/c, &$E_{\rm{lab}}> 500$ MeV,\nonumber
    \end{subnumcases}

\begin{figure*}[!hpbt]
    \includegraphics[width=0.3\textwidth]{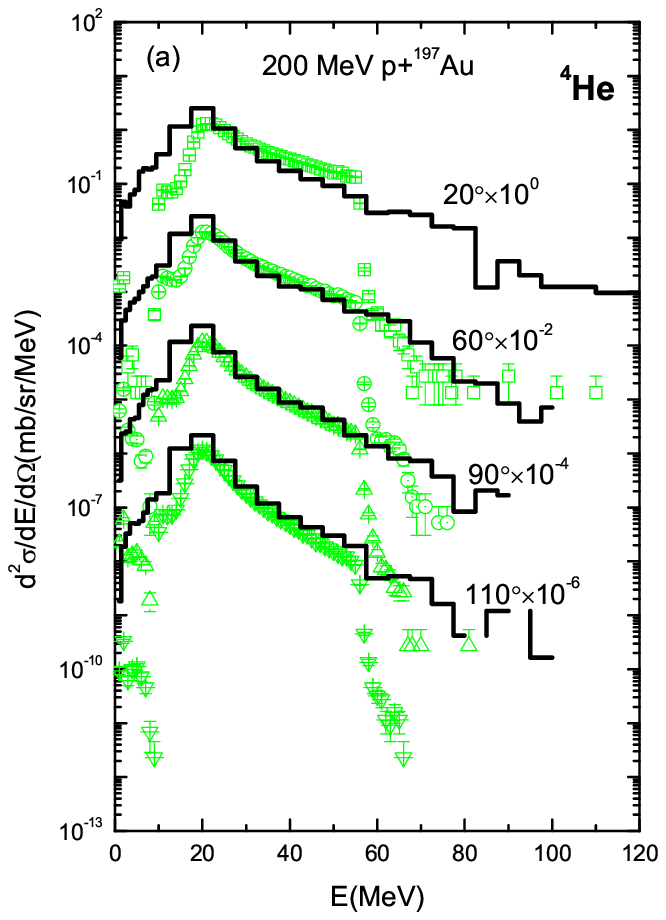}
    \includegraphics[width=0.3\textwidth]{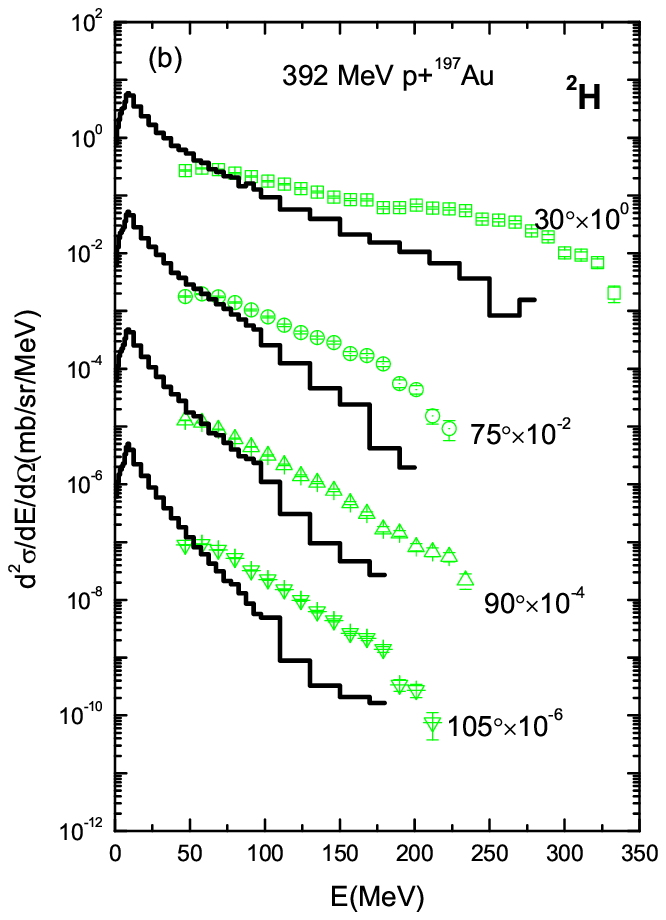}
    \includegraphics[width=0.3\textwidth]{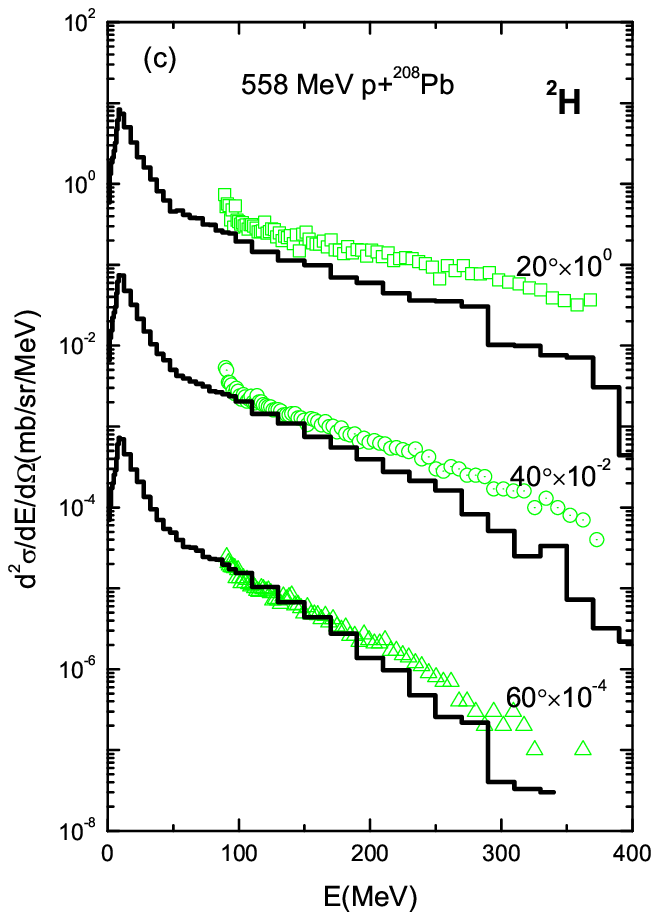}\\
    \includegraphics[width=0.3\textwidth]{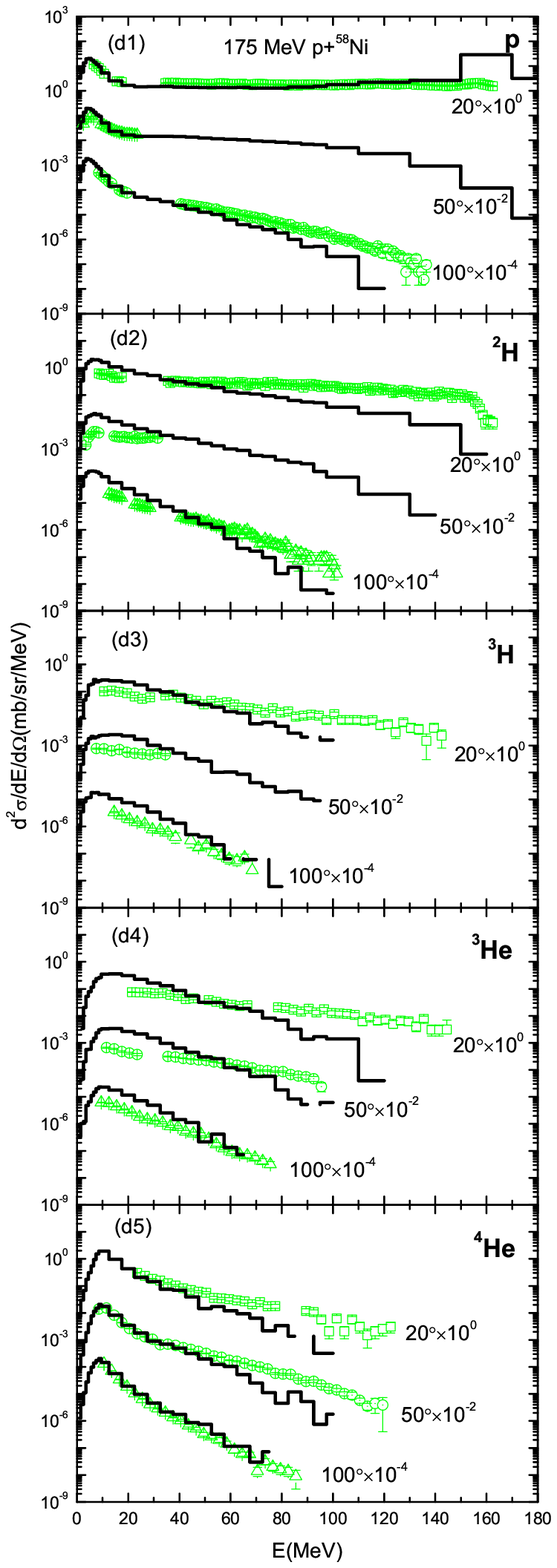}
    \includegraphics[width=0.3\textwidth]{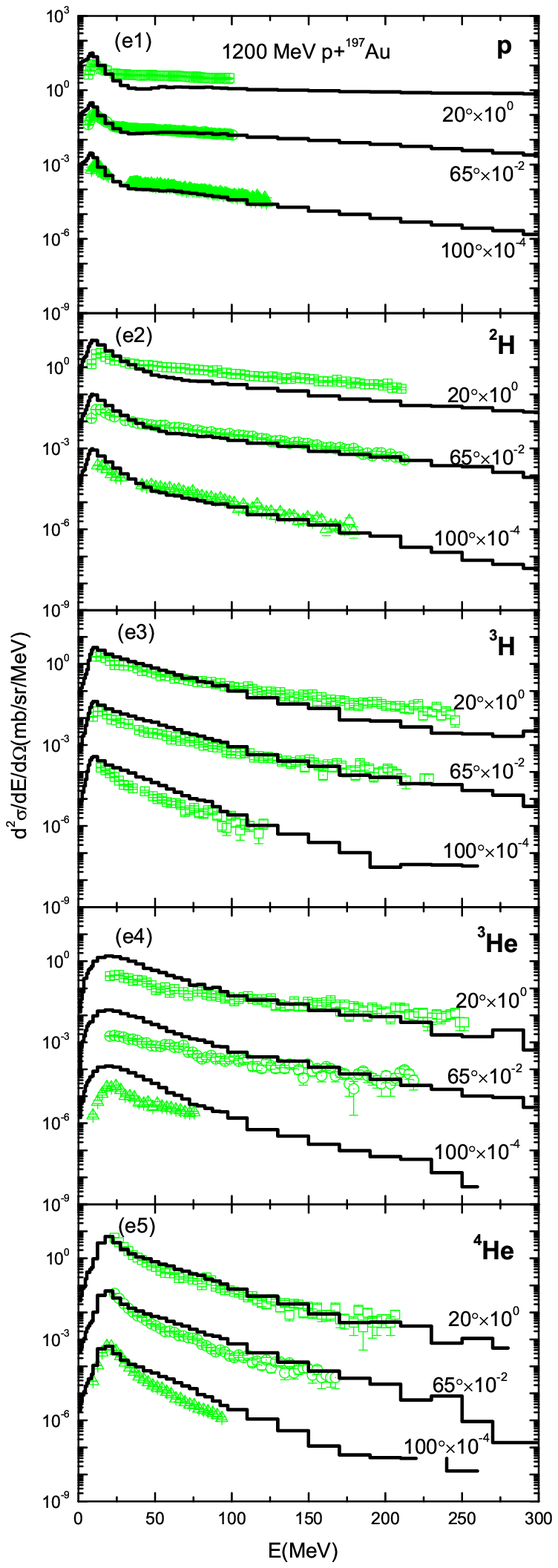}
    \includegraphics[width=0.3\textwidth]{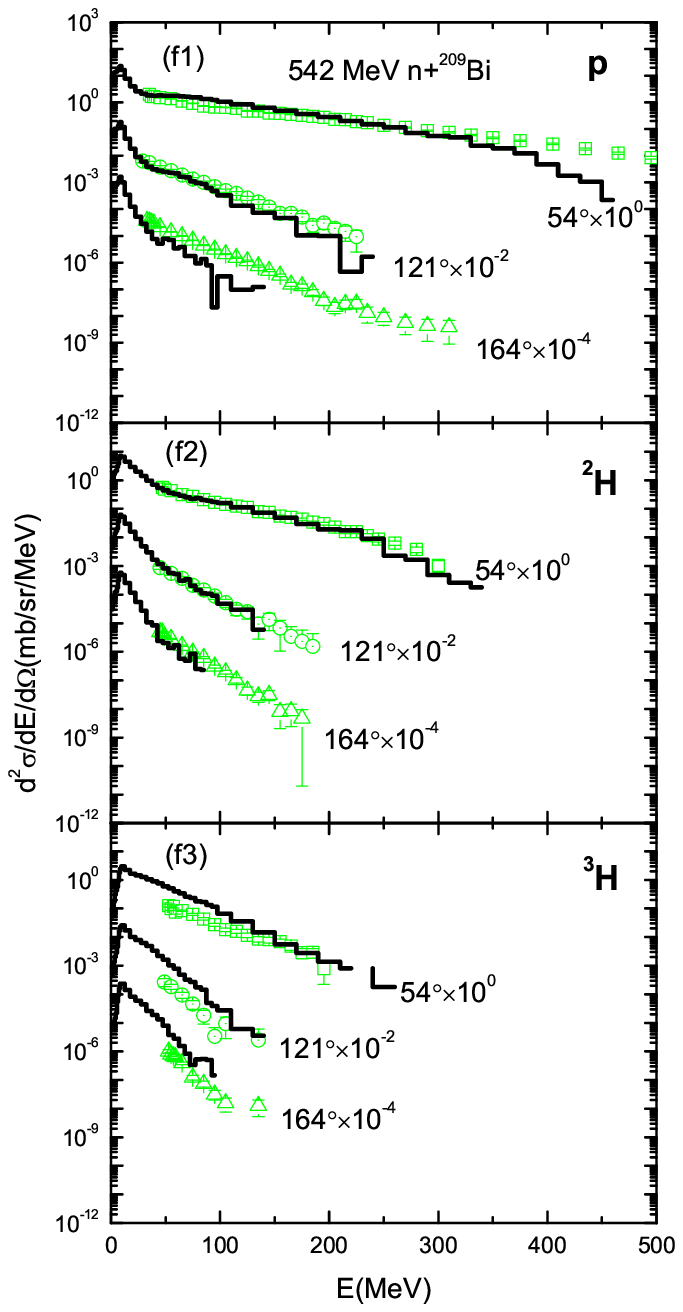}
    \caption{(Color online) Calculated DDXs of LCPs
    produced in the reactions of nucleon bombarding various targets at energies from 175 to 1200 MeV.
    Experimental data are taken from Ref.\cite{Mach1} for 200 MeV $p+^{197}$Au,
    Ref.\cite{Uozumi2011} for 392 MeV $p+^{197}$Au, Ref.\cite{Beck2} for 558 MeV $p+^{{\rm nat}}$Pb,
    Ref.\cite{Piskor} for 175 MeV $p+^{58}$Ni,
    Ref.\cite{Budza} for 1200 MeV $p+^{197}$Au, and Ref.\cite{Franz} for 542 MeV $n+^{209}$Bi,
    respectively.}
    \label{fig6}
    \end{figure*}

    To test the prediction power of model, with the fixed parameters in surface coalescence model,
    more reactions with various incident energies
    and targets are simulated systematically. The comparisons between calculated DDXs of LCPs and experimental data
    for 175 MeV proton hitting $^{58}$Ni, 200, 392, 1200 MeV proton hitting $^{197}$Au,
    542 MeV neutron hitting $^{209}$Bi, and 558 MeV proton hitting $^{208}$Pb, are illustrated in Fig.~\ref{fig6}.
    From the figure, one can see that the calculations are overall in fairish agreement with experimental data.
    It means that present ImQMD model has ability to describe the nuclear data of free nucleons and
    LCPs, including $d$, $t$, $^3$He, and $^4$He, with a uniform code.

\section{summary}

    In order to overcome the limitation of ImQMD05 model in description on LCPs emission,
    a phenomenological surface coalescence mechanism is introduced into ImQMD05 model.
    The base idea of this mechanism is that: The leading nucleon
    ready to leave from compound nuclei can coalesce with other nucleon(s) to form a LCP,
    and those LCPs with enough kinetic energies to overcome Coulomb barrier can be emitted.
    By systematic comparison between calculation results and
    experimental data of nucleon-induced reactions,
    the parameters in the surface coalescence model are fixed. Then with the fixed parameters,
    chosen once for all, the prediction power of the model is tested
    by the nucleon-induced reactions on various targets with energies from 62 to 1200 MeV.
    And it is found that, with surface coalescence mechanism introduced into ImQMD model,
    the description on the DDXs of LCPs is great improved.

    \begin{acknowledgments}
    This work was supported by the National Natural Science Foundation
    of China under Grant Nos. 11005022, 
    11365004, 
    11365005, 
    11075215, 
    11005003, 
    11275052, 
    and by the Doctor Startup Foundation of Guangxi Normal University.

    \end{acknowledgments}




\end{document}